\documentclass[aps,prl,showpacs,twocolumn,
preprintnumbers,amsmath,amssymb]{revtex4}

\usepackage{graphicx}
\usepackage{amsmath,amssymb}
\usepackage{bm}
\usepackage{ifthen}

\renewcommand{\vec}[1]{{\bm #1}}

\begin{document}

\title{Hidden correlations of exciton complexes in self-assembled
  quantum dots and rings}
\author{Weiwei Zhang}
\author{Ming Gong}
\author{Chuan-Feng Li}
\author{Guang-Can Guo}
\author{Lixin He\footnote{corresponding author: helx@ustc.edu.cn}}
\affiliation{Key Laboratory of Quantum Information,
University of Science and Technology of China,
Hefei, 230026, People's Republic of
China}
\date{\today}

\begin{abstract}

The binding energies of trions ($X^+$, $X^-$) and biexciton (XX)
in self-assembled
semiconductor quantum dots (QDs) are very sensitive to the
geometry and chemical
composition of the QDs, and are random from dots to dots.
However, in this letter, we show through analytical and
numerical methods that
the transition energies of the exciton complexes
in self-assembled
quantum dots and rings follow a simple and robust
rule, i.e., the sum of exciton and biexciton transition energies minus the
transition energies of trions
is always positive and almost
a constant for the same type of quantum dots and rings
as a consequence of a pure Coulomb correlation effect.
More interestingly, this quantity
show a sharp transition when the topology change from 
a dot to a ring.
This {\it hidden} correlation effect,
directly measurable in experiments, offers a
useful way to understand the
photoluminescence spectra of self-assembled
quantum dots and rings.

\end{abstract}



\maketitle

New physics emerge when the size of
a solid system reduces to nano-scale.
In the
self-assembled semiconductor quantum dots (QDs),
the three-dimensional confinement effects
lead to atom-like electronic structure and
long-living coherent quantum states.
The confinement also enhances the Coulomb interactions among the
quasi-particles, leading to novel physics in QDs other
than bulk materials, e.g., Coulomb blockade effects\cite{field93}, and the
non-Aufubau filling order for holes \cite{reuter05,he05d}
in the InAs/GaAs quantum dots. The unique
properties of QDs are not only of special interests in the view of
fundamental physics, but also have important applications in quantum
information processes.
In these applications, the exciton complexes, including exciton
($X$), biexciton ($XX$) and trions ($X^+$, $X^-$) play extremely
important roles.
For example, neutral excitons can be used to
generate single-photons \cite{michler00, yuan02}, whereas biexcitons
can be used to generate entangled photon pairs \cite{akopian06,
stevenson06}.
Trions can be used to write in/read out the information of spin
qubit, or to manipulate the spin states \cite{berezovsky06,
berezovsky08, press08, gerardot08}.
Due to the enhanced Coulomb interactions, the transition energies of
the biexcitons and trions have significant (a few meV) energy shifts
(i.e., binding energy) relative to those of the neutral excitons.
Question arise that if we can find some simple relations between the
transition energies of these exciton complexes?

Unfortunately, both experimental measurements \cite{rodt05,
seguin06} and theoretical calculations
\cite{bester03b} show that the binding energies
of the exciton complexes change dramatically, even their signs, with
respect to the sizes, shapes and chemical compositions of the QDs.
Even worse, the binding energies seems to be random from dot to dot.
It seems hopeless to find some simple relations between the
transition energies of the exciton complexes. In this letter, we
show however, through analytical analysis and numerical calculations
that there is indeed a simple and robust relations among the
transition energies of the exciton complexes in self-assembled
quantum dots and rings.

The binding energies of exciton complexes vary mainly due to the
competition of the direct electron-electron, hole-hole and electron-hole
Coulomb interactions\cite{bester03b, rodt05},
which strongly depend on the size or the composition of
a QD.
To eliminates the influence of direct Coulomb interactions,
we define a quantity
$\Delta$,
\begin{equation}
\Delta=X+XX-X^{+}-X^{-} \, ,
\label{eq:definition1}
\end{equation}
where, $X$, $XX$, $X^{+}$, and $X^{-}$ are the transition
energies of excitons, biexcitons, positive and negative
trions, respectively. It is easy to show that $\Delta$=0
under the Hartree-Fock (or single configuration) approximation,
as the direct Coulomb energies and exchange energies
cancel each other for the four types of exciton complexes.
However, $\Delta$ is generally not zero when 
we include correlation energies.

We have calculated the  photoluminescence (PL) spectra for
a large amount of QDs with different sizes
(radii and heights),
compositions (InGaAs/GaAs or InPAs/InP)
and shapes (lens, cone, elongated and pyramidal dots).
We first obtain the single particle energy levels and wave functions
of the geometry-optimized QDs using an empirical pseudopotential
method \cite{williamson00}, where the total pseudopotential of the
system are superposition of the local, screened atomic
pseudopotential of all (dot+matrix) atoms and the nonlocal
spin-orbit potentials. The pseudopotential Schr\"{o}dinger equation
is solved via the linear combination of bulk bands (LCBB)
method~\cite{wang99b}. Many-body effects are included via the
configuration interaction (CI) method \cite{franceschetti99} by
expanding the total wavefunction in Slater determinants for single
and bi-excitons formed from all of the confined single-particle
electron and hole states.

\begin{figure}
\begin{center}
\includegraphics[width=2.8in,angle=-90]{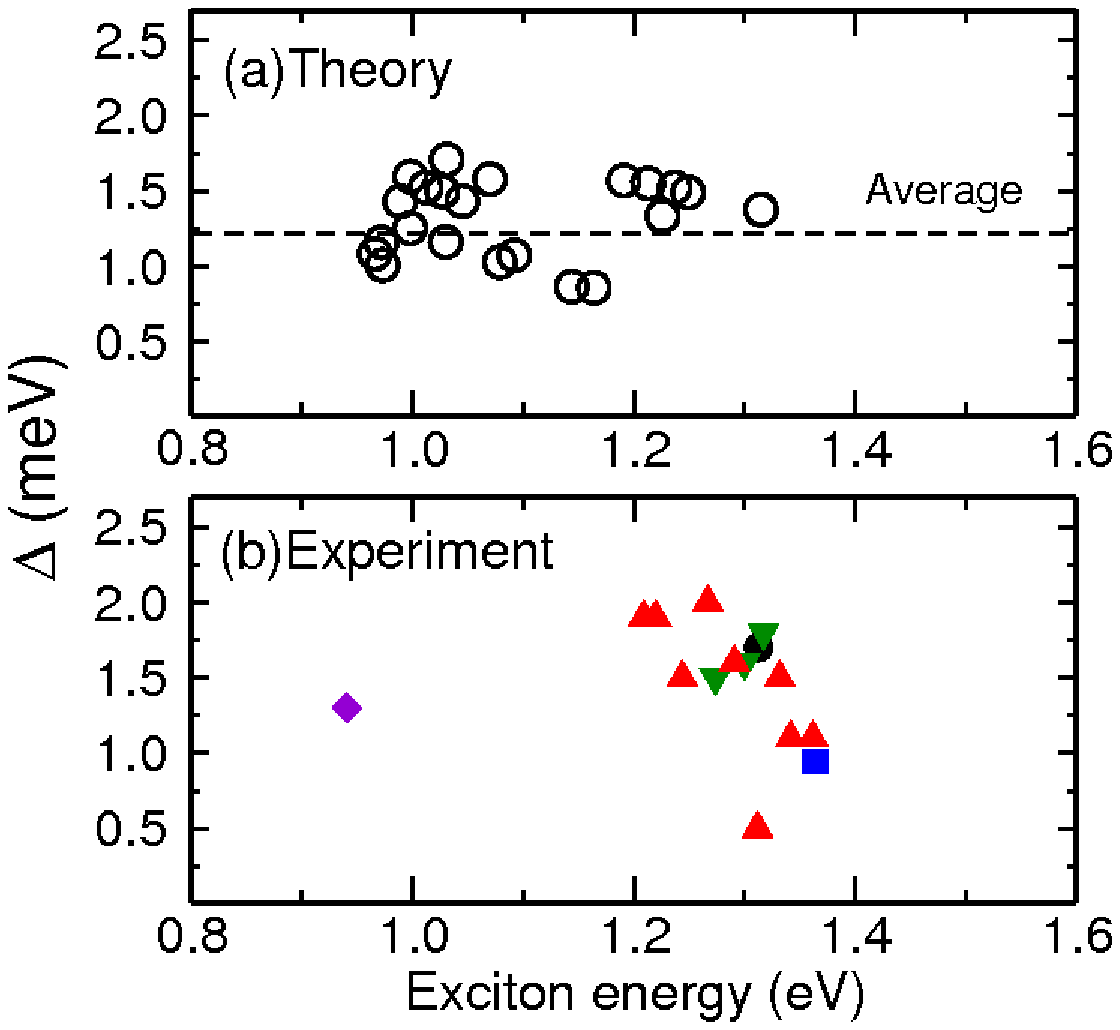}
\end{center}
\caption{ (color online) (a) Calculated $\Delta$ for InGaAs/GaAs QDs
with various sizes and geometries. (b) Experimental $\Delta$
extracted from literatures. ($\blacktriangle$), ($\blacktriangledown$)
($\bullet$), ($\blacklozenge$), and ($\blacksquare$) are
extracted from  Ref. \onlinecite{rodt05}, 
\onlinecite{seguin06}, \onlinecite{dalgarno06}, 
\onlinecite{cade05}, and \onlinecite{shields07}, 
respectively.}
\label{fig:theory}
\end{figure}

Figure \ref{fig:theory}(a) depicts the calculated $\Delta$ in
InGaAs/GaAs QDs with respect to their exciton energies. It clearly
shows that $\Delta$, in contrast to the binding energies, are always
positive and distributed in a narrow energy range from 0.8 to 1.8
meV even when the exciton energy change dramatically from 0.9 to
1.4 eV, with an average value equal to about 1.22 meV. We see
that $\Delta$ hardly depends on the size, composition, and even the
shape of the QDs. Therefore $\Delta$ represent a simple and robust
relationship between the transition energies of exciton complexes.
To verify the calculations, we show in Fig. \ref{fig:theory}(b)
experimental values of $\Delta$ extracted from the InGaAs/GaAs QDs
PL spectra in the literatures\cite{dalgarno06, cade05, seguin06,
shields07, rodt05} of four different groups. Indeed, the
experimental $\Delta$ distributes in 1.0 - 2.0 meV as the exciton
energy varies from 0.9 to 1.4 eV, which strongly support the
theoretical predictions. We have also calculated the $\Delta$
for InAs/InP QDs and find that $\Delta$ are distributed in 0.6 - 1.3
meV.

Vanishes in the HF approximation, $\Delta$ is therefore
a pure correlation effect. It not only represents
a simple relation between the transition energies of
the excition complexes, but also
provides an easy way to characterize
the correlation effects in different
nano-structures, which was previously only
avaliable in theoretical calculations.
In order to gain an analytic understanding of these
results, we take use of a perturbation theory.
We use a typical lens-shaped InAs/GaAs QD as an example.
Other kind of dots can be studied in a similar way.
Since  $\Delta$=0 under Hartree-Fock approximation,
$\Delta$ can be alternatively calculated as,
\begin{equation}
\Delta=\Delta_{XX} - \Delta_{X^{+}} - \Delta_{X^{-}} \, ,
\label{eq:definition2}
\end{equation}
where, $\Delta_{X^{+}}$, $\Delta_{X^{-}}$, and $\Delta_{XX}$ are
the correlation energies (i.e., the energy difference between CI and HF calculations)
for the trions and biexcitions, respectively.
For example,
$\Delta_{X^+}=E_{\rm CI}(X^+)-E_{\rm HF}(X^+)$.
To get the correlation energies for each type of excitons,
we first solve the many-particle
Hamiltonian in a single configuration
approximation
and obtain the energy
$E_{j}^{(\rm {\rm HF})}$ for $j$-th configuration $|\Phi_{j}\rangle$.
To the second order approximation,
the correlation energy of an exciton is,
\begin{equation}
\Delta_{X} =  \sum_{j \neq 0} {|\langle \Phi_{0} (X) |H_{I}|\Phi_{j} (X)
\rangle|^{2} \over E_{0}^{(\rm{HF})}(X)-E_{j}^{(\rm{HF})}(X)}
\label{eq:correlation}
\end{equation}
where, $H_I$ is the Coulomb interactions among the quasi-particles.
$|\Phi_{0} (X) \rangle$ and $E_0^{(\rm{HF})} (X)$
are the ground state configuration and its
energy for an exciton.
The correlation energies of $XX$,and
trions $X^+$, $X^-$ can be calculated in the exact same procedures.

\begin{figure}
\begin{center}
\includegraphics[width=2.8in]{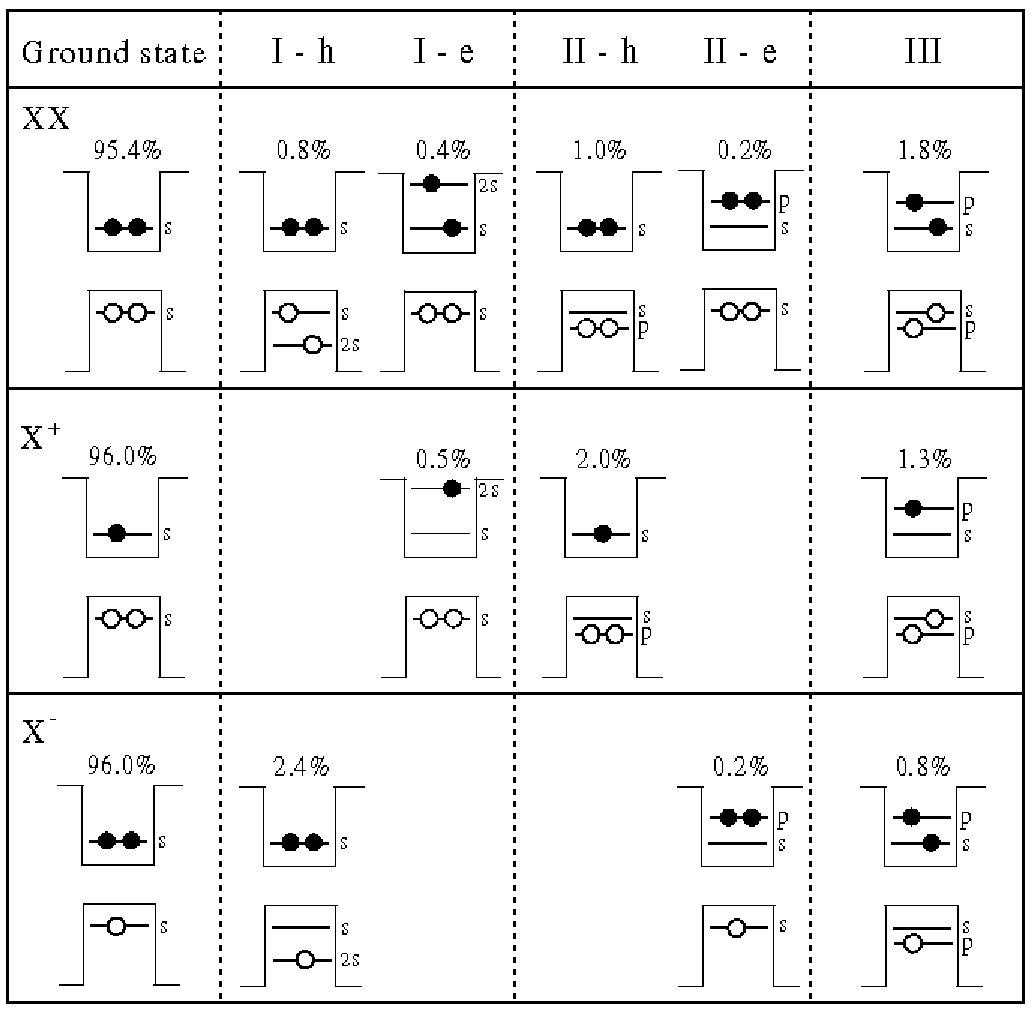}
\end{center}
\caption{A schematic figure of
the excited configurations that couple substantially to the ground
  state configurations for $XX$, $X^+$ and $X^-$,
divided in three categories (see text for detail). The numbers of
percentage in each configuration is the weight of this configuration
in the total many-particle wavefunctions from CI calculations.
We omit irrelevant single particle levels. }
\label{fig:excitation}
\end{figure}

We show schematically in Fig. \ref{fig:excitation} the excited configurations that
couple substantially to the
ground states of $XX$, $X^{+}$ and $X^{-}$.
These configurations can be classified into three categories.
We define the configurations in which only one electron (or hole) is
excited as type I configurations.
Type II configurations are those that two electrons (or two holes)
are simultaneously excited, whereas
type III configurations contain one excited electron and one
excited hole.
Configurations of more than two particle excitations do not couple to the
ground state.

Type I configurations can be further divided as electron excitations (type I-e) and
hole excitations (type I-h).
As we see from Fig.~\ref{fig:excitation},
for $X^{+}$ only type I-e excitations contribute significantly to the
correlation energy, whereas for $X^{-}$, only type I-h excitations
contribute significantly.
On the other hand,  both I-e and I-h contribute to the correlation energies of
biexcitons.
The off-diagonal (OD) matrix elements of I-h excitation for $XX$ is
$(H_{I-h})_{XX}=2 \langle e_{s},h_{s};e_{s},h_{2s} \rangle
+ \langle h_{s},h_{s};h_{s},h_{2s} \rangle$
and for $X^{-}$ is $(H_{I-h})_{X^{-}}=2 \langle e_{s},h_{s};e_{s},h_{2s} \rangle$.
Here, we used a short notation,
$\langle e_{s},h_{s};e_{s},h_{2s} \rangle =$
\begin{equation}
\int\int\frac{\phi^{*}_{e_{s}}(\vec{r}_{1})\phi^{*}_{h_{s}}
(\vec{r}_{2})\phi_{e_{s}}(\vec{r}_{1})
\phi_{h_{2s}}(\vec{r}_{2})}{\epsilon ({\bf r}_1 -{\bf r}_2)
|\vec{r}_{1}-\vec{r}_{2}|}d^3
\vec{r}_{1}d^3\vec{r}_{2}
\, .
\end{equation}
where, $\phi_{e_{s}}(\vec{r}_{1})$, $\phi_{h_{s}}(\vec{r}_{2})$,
$\phi_{h_{2s}}(\vec{r}_{2})$ are the electron ($e$) and
hole ($h$) single particle wavefunctions, and
$\epsilon ({\bf r}_1 -{\bf r}_2)$ is the dielectric function 
\cite{franceschetti99}.
For simplicity, we dropped the spin notation in the above equation, but
included it in real calculations.
Since electron and hole wavefunctions have similar
dimensions in QDs
\cite{williamson00}, $\langle e_{s},h_{s};e_{s},h_{2s} \rangle$ $\sim$
-$\langle h_{s},h_{s};h_{s},h_{2s} \rangle$. Therefore
$ \langle h_{s},h_{s};h_{s},h_{2s} \rangle$ almost cancels
$ \langle e_{s},h_{s};e_{s},h_{2s} \rangle$.
As a result,
$(H_{I-h})_{XX}$ is much smaller than
$(H_{I-h})_{X^{-}}$.
This cancellation effect is also the reason that I-e excitations
contribute little to the correlation
energy of $X^+$.
For the other I-h configurations,
the OD term vanishes in cylindrical dots ,
because
$\langle e_{s},h_{s};e_{s},h_{p} \rangle$,
$ \langle e_{s},h_{s};e_{s},h_{d} \rangle$=0
due to the symmetry of the wavefunctions.

The difference of the denominators of in calculating the correlation energies
of $XX$ and $X^-$ [see
Eq. (\ref{eq:correlation})] also contributes to $\Delta$.
The single-particle part of the denominators are the same for
$XX$ and $X^-$, and
the difference comes from many-particle interactions.
It turns out that for $XX$ and $X^-$,
the difference between the two denominators
is $J^{hh}_{ss}-J^{hh}_{s-2s}$, i.e., the difference of Coulomb energies between the
hole $s$ state with itself and $s$ with 2$s$ states.
This energy difference is much smaller than the denominators
themselves.
The correlation energy of $XX$ from I-h excitations has
a additional factor of 2 due to spin degeneracy.
Combining all above factors, we find that I-h contribution to the correlation energies
$|\Delta_{X^{-}}^{I-h}|$ is much larger than
$|\Delta_{XX}^{I-h}|$.
Using the numerically calculated Coulomb
integrals and single particle levels, for a lens-shaped
dot with base $D$=25 nm, height $h$=2 nm,
$\Delta_{X^{-}}^{I-h}$$\approx$ 2.5 $\Delta_{XX}^{I-h}$.
The I-h configurations' contribution to $\Delta$ is about 0.93 meV.

We can apply similar analysis to type I-e configurations,
and find they also make positive contribution
to $\Delta$. However, the contribution from I-e excitations
are much smaller than those of
I-h ones, because
electrons have much larger level spacing than that of holes
in the InAs/GaAs dots and therefore larger
denominators (about 3 times) in Eq. \ref{eq:correlation}  than
those I-h excitations.
For the same dot above, I-e configurations contribute 0.31 meV to $\Delta$.

Type II configurations can also be sub-classified into II-h and II-e excitations.
The II-h OD matrix elements for
$XX$ are $(H_{II-h})_{XX}=\langle h_{s},h_{s};h_{i},h_{j} \rangle$,
where $i$, $j$ are the indices of the excited single-particle levels.
If $i \neq j$, the configurations
would not couple to the ground state configuration
in cylidrical dots. For $i$=$j$ configurations,
the most contributions are from $p$ orbitals excitions.
The contributions from higher
energy excitions are much smaller due to larger energy spacing.
Interestingly, we find that
$X^{+}$ has the same OD matrix elements for II-h excitations
i.e. $(H^{OD}_{II-h})_{X^+}=(H^{OD}_{II-h})_{XX}$.
As a result, only the difference between the denominators 
contributes to $\Delta$.
For type II-h configurations,
the difference between two denominators $\sim$
2$(J^{eh}_{ss} - J^{eh}_{si})$.
We find that $\Delta
E_{II-h}^{\rm HF} ({XX})$ is always larger than $\Delta
E_{II-h}^{\rm HF} ({X^{+}})$, because the coulomb integrals always
satisfy $J^{eh}_{ss} > J^{eh}_{si}$. As a result, type II-h
configurations always make positive contribution to $\Delta$,
but usually much smaller than those of type I configurations.
Similar analysis applies for
type II-e configurations. The denominator
$\Delta E_{II-e}^{\rm HF} ({XX}) > \Delta
E_{II-e}^{\rm HF} ({X^{-}})$.
Therefore, type II-e configurations also make
positive contribution to $\Delta$. However, due to much larger
single particle level spacing of electrons,
the contribution from type II-e
configurations is even smaller than that from type II-h ones.

Type III configurations contain one excited electron and one excited
hole, as shown in Fig.~\ref{fig:excitation}. We find that the OD
Hamiltonian matrix elements for for $XX$, $X^{+}$ and $X^{-}$ are
the same for each type of excitation, i.e.,
$(H^{OD}_{III})_{XX}=(H^{OD}_{III})_{X^{+}}=
(H^{OD}_{III})_{X^{-}}=\langle e_{s},h_{s};e_{i},h_{j} \rangle$,
where $i$, $j$ are the indeces of the excited single particle levels
in the configurations. The denominators in calculating the
correlation energies for these three types of exciton complexes 
[see Eq. (\ref{eq:correlation})]
are also very close for type III configurations. 
After consider the spin degeneracy, 
the correlation energy of $XX$ almost cancels
those of trions, and type III configurations have very little
contribution to $\Delta$.

For typical cylindrical dots,
the hidden correlation  $\Delta$ mostly comes from
type I configurations. The type II contribution is less but
significant, whereas the type III contribution is negligible.
For example, for the lens-shaped dot with $D$=25 nm and $h$=2 nm,
the perturbation theory gives $\Delta$=1.51 meV, in which
type I configurations contribute 1.23 meV and type II ones contribute
the remainder 0.28 meV. As to the individual configurations, the type I
hole 2$s$ excitation have the most significant contributions to $\Delta$.
This non-trivial result suggests that not including the 2$s$ orbital in the CI
basis will cause huge error in calculating $\Delta$.
The perturbation theory gives a little smaller $\Delta$ than that from
CI method (1.80 meV), due to ignoring higher order terms.
We have applied the same analysis for other type dots, with
different sizes and shapes (e.g. elongated dots and pyramidal dots).  
Even though the
details of the contributions from each configuration may be quite different,
the final results of $\Delta$ are very close, which suggests that there are 
more fundamental physics remain to be resolved in future studies.  

\begin{figure}
\begin{center}
\includegraphics[width=3.0in]{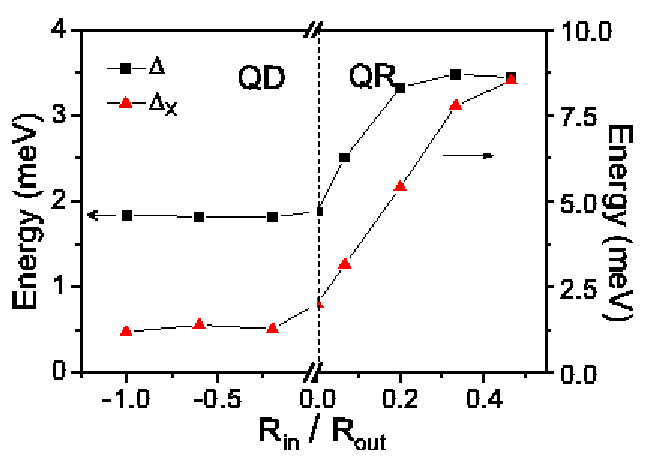}
\end{center}
\caption{The correlation energy of the exciton $\Delta_X$ and the
{\it hidden}
  correlation $\Delta$ as functions of $R_{\rm in} /R_{\rm out}$ in a
InAs/GaAs nanostructure. When $R_{\rm in} /R_{\rm out} <$0, the
nanostructure is a QD, and becomes a QR when $R_{\rm in} /R_{\rm
out}
>$0. } \label{fig:delta}
\end{figure}

We now discuss another type of important semiconductor
nano-structures, quantum rings (QRs), 
which have non-simply connected topology.
In a QR, the single-particle energy level is more sensitive to the
inner radii of the ring, whereas the Coulomb interactions dependent
more strongly on the outer radii of the rings \cite{li01b, zhang08}.
The single-particle energy levels in a QR are much more condensed
than in a QD \cite{li01b,zhang08} of similar sizes, and therefore
the particles are expected to be more correlated.
Figure \ref{fig:delta} depicts the correlation energy of an exciton
$\Delta_{X}$ and $\Delta$ with respect to the ratio of inner
radius to outer radius ($R_{in}/R_{out}$) in a InAs/GaAs QR.
We adapt the ring model from Ref. \onlinecite{zhang08}.
The outer radius of the ring is fixed at $R_{out}$= 15 nm, whereas
the inner radius $R_{in}$ increase from -15 nm to 7 nm.
The structure with $R_{in} <$ 0 is a QD and the structure with $R_{in} > 0$
is a QR.
Figure \ref{fig:delta} shows that when the structure changes from a QD to a
QR, $\Delta_{X}$ increases dramatically from about 2 meV to about 10 meV.
Remarkably, almost a constant in the QD region, $\Delta$ 
shows a sharp transition when the topology changes
from a dot to a ring, where $\Delta$ jumps 
from 1.8 meV in the QD region to 3.5 meV in
the QR region and
saturates when $R_{in}/R_{out}$ exceeds 0.2.
This effect can be directly measured in a dot to ring transition experiments
\cite{granados03}.

To conclude, we have introduced a {\it hidden} correlation function
as the sum of exciton and biexciton transition energies minus the
transition energies of trions
for the exciton complexes in semiconductor nanostructures
such as quantum dots and rings. Measurable in experiments, the
hidden correlation provides a deep insight to
the Coulomb correlation effects of the exciton complexes.
We show that the hidden correlation energy 
is positive and almost a constant for the same type of quantum dots. 
Remarkably, it shows a sharp 
phase-transition-like behave when the topology changes from
a dot to a ring. 
The fundamental physics that govern this intriguing behaver 
remain to be fully explored. 

L.H. acknowledges the support from the Chinese National
Fundamental Research Program 2006CB921900, the Innovation
funds and ``Hundreds of Talents'' program from Chinese Academy of
Sciences, and National Natural Science Foundation of China (Grant
No. 10674124).


\end{document}